\pdfoutput=1

\documentclass[11pt]{article}
\usepackage[table]{xcolor}
\usepackage[preprint]{acl}
\usepackage{graphicx}
\usepackage{times}
\usepackage{latexsym}
\usepackage[T1]{fontenc}
\usepackage[utf8]{inputenc}
\usepackage{microtype}
\usepackage{booktabs}
\usepackage{amsmath}
\usepackage{multirow}
\usepackage{array}
\usepackage{tabularx}
\usepackage{tikz}
\usepackage{dblfloatfix}
\usepackage{float}
\usepackage{placeins}
\usetikzlibrary{arrows.meta,positioning}
\usepackage{aimsheader}  % AIMS Lab corner logo (arXiv preprint only)
\hypersetup{
  pdftitle={Same Ranking, Different Winner: How Scoring Targets Shape LLM Memory Benchmarks},
  pdfauthor={Sugam Panthi, Rabab Abdelfattah}
}

\definecolor{RawTurnColor}{HTML}{4C72B0}
\definecolor{SourceFamilyColor}{HTML}{55A868}
\definecolor{CanonicalColor}{HTML}{C44E52}
\colorlet{TableHeaderColor}{black!8}
\newcommand{\theadrow}{\rowcolor{TableHeaderColor}}
\newcommand{\rawcell}[1]{\cellcolor{RawTurnColor!18}{#1}}
\newcommand{\sourcecell}[1]{\cellcolor{SourceFamilyColor!18}{#1}}
\newcommand{\canonicalcell}[1]{\cellcolor{CanonicalColor!18}{#1}}
\newcommand{\rawtxt}[1]{\textcolor{RawTurnColor!85!black}{#1}}
\newcommand{\sourcetxt}[1]{\textcolor{SourceFamilyColor!85!black}{#1}}
\newcommand{\cantxt}[1]{\textcolor{CanonicalColor!85!black}{#1}}
\newcommand{\mtelmem}{MTEL-Mem}
\newcommand{\pairRS}{\cellcolor{TableHeaderColor}\rawtxt{Raw} vs \sourcetxt{Source}}
\newcommand{\pairRC}{\cellcolor{TableHeaderColor}\rawtxt{Raw} vs \cantxt{Canonical}}
\newcommand{\pairSC}{\cellcolor{TableHeaderColor}\sourcetxt{Source} vs \cantxt{Canonical}}
\newcommand{\figref}[1]{\hyperref[#1]{Figure~\ref*{#1}}}
\newcommand{\tabref}[1]{\hyperref[#1]{Table~\ref*{#1}}}

\title{Same Ranking, Different Winner: How Scoring Targets Shape LLM Memory Benchmarks}

\author{Sugam Panthi \quad Rabab Abdelfattah \\
  The University of Southern Mississippi \\
  \texttt{\{sugam.panthi, rabab.abdelfattah\}@usm.edu}}

\begin{document}
\aimscornerlogo  % AIMS Lab logo, top-left of page 1
\maketitle

\begin{abstract}
Conversational-memory systems increasingly transform dialogue history
into facts, summaries, timelines, and other source-linked descendants,
so a single source turn can coexist with several derived memories in the
same retrieval index. This raises an under-specified evaluation question:
which stored form should receive retrieval credit? We show that this
\emph{scoring-target} choice is often left implicit and can materially
change benchmark conclusions. We present \textsc{TIAP}, a fixed-output
audit that rescores saved ranked outputs under three targets---Raw,
Source, and Canonical---without rerunning retrieval. On LoCoMo and
LongMemEval-S, switching only the credited target changes nDCG on
83.4\%--94.0\% of shared queries, flips target orderings on Mem0 and
MemoryOS transfer runs, and reverses parser-density recommendations.
A 1{,}902-case semantic audit further shows that relaxed source-linked
credit is fully justified only 29.2\% of the time, despite high rubric
reliability in a validation subset. These results reveal
\emph{target noninvariance}: conclusions about memory architectures can
silently flip with a single benchmark-design choice. Conversational-memory
papers should therefore define and report the scoring target explicitly.\footnote{\href{https://github.com/rabdelfattahlab/LLM_Memory_Benchmark}{[GitHub]}}
\end{abstract}

\section{Introduction}

\begin{figure}[!t]
 \centering
 \includegraphics[width=\columnwidth]{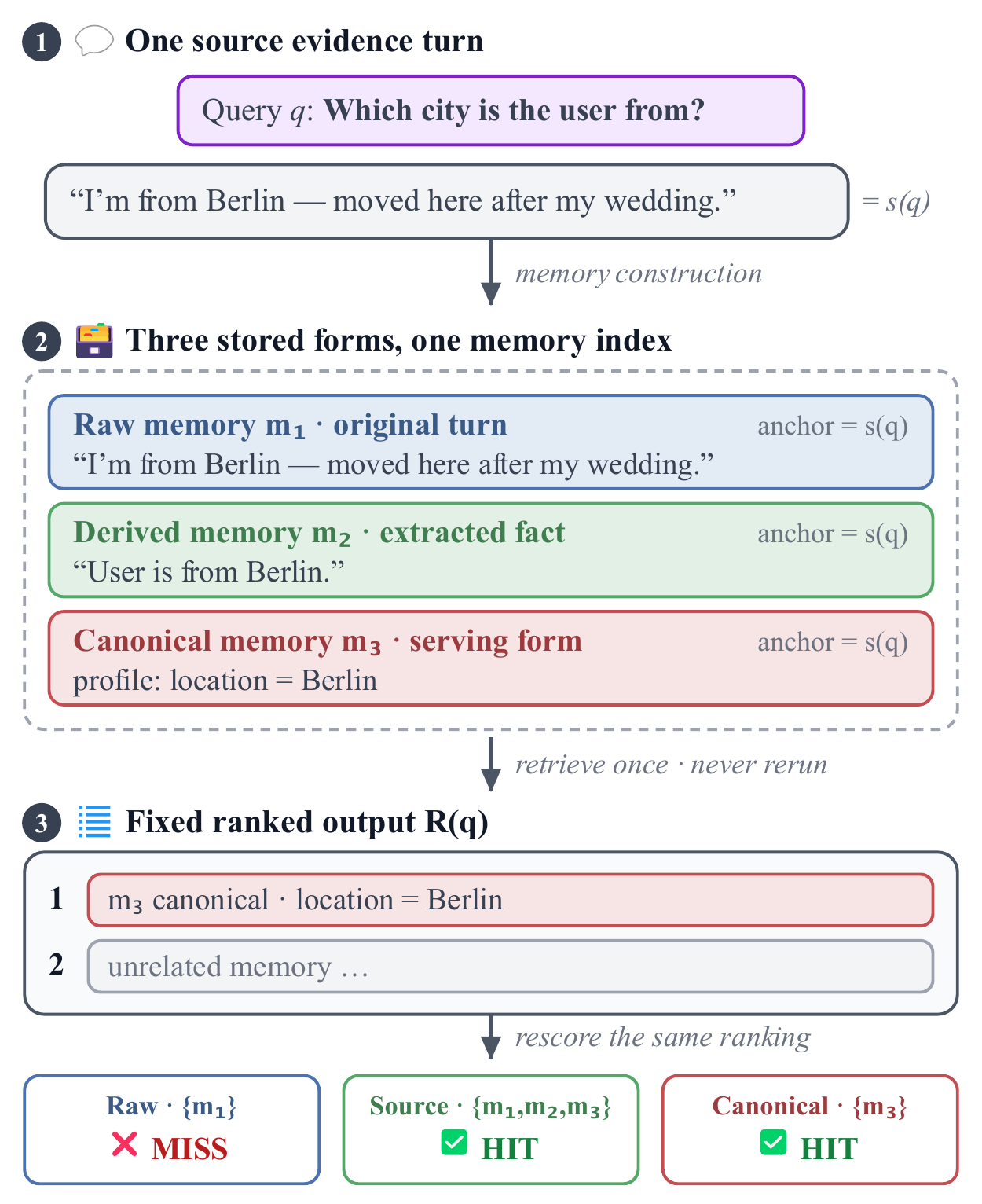}
 \caption{
 Overview of target noninvariance in conversational memory evaluation.
 A single source evidence turn can produce multiple stored memory forms
 in the same index. For a fixed ranked output, changing only the
 credited target can turn the same retrieval result into a miss or a hit.
 }
 \label{fig:overview}
 \vspace{-4 mm}
\end{figure}

Long term conversational AI systems are increasingly built around
\emph{memory construction}. Rather than passing raw dialogue history
directly to the response model, modern chat assistants extract facts,
consolidate updates, build timelines, populate profile memories, and
often retain the original dialogue turn alongside its transformed
descendants in the same retrieval index \citep{mem0, memoryos, zep,
theanine, rmm}. This storage pattern is now central to memory-augmented
agents \citep{premem, theanine, longmemeval}, but it also creates an
under-specified evaluation problem. A benchmark query is typically
anchored to one source turn, while the evaluated store may contain
several retrieval units derived from that same evidence. Which of these
stored forms should receive retrieval credit?

Figure~\ref{fig:overview} summarizes the core failure mode. A single
source evidence turn may be stored as a raw memory, a source-linked
derived memory, and a canonical serving memory. When the ranked output
is fixed, the same retrieved item can be scored as a miss under a Raw
target but as a hit under Source or Canonical targets. Thus, the
benchmark conclusion can change even though retrieval behavior has not
changed.
The ambiguity has both semantic and numerical consequences. Semantically, a stored
descendant may share the correct source anchor while omitting the answer,
dropping an important condition, or preserving only weakly related
context. In our 1{,}902 case audit of relaxed credit, only 29.2\% of
credited source-linked descendants fully support the query under a
five-model majority vote. Thus, a more permissive target is not
automatically a more correct one, and answer support can only be judged
after the benchmark decides which stored evidence units are admissible
\citep{ragas, ares}.

Numerically, target choice is not a bookkeeping artifact. If relaxing
the target merely raised recall-like scores, the broadest target would
dominate. Instead, target choice interacts with benchmark structure and
memory architecture. In transfer runs, target orderings are non-monotonic for both Mem0 and MemoryOS, depending on the benchmark. Thus, the same fixed ranked outputs can
recommend different memory-store designs depending only on which stored
representation is credited.
Because conversational-memory benchmarks increasingly guide retriever, storage-policy, and serving-representation choices, comparisons are only meaningful if the benchmark defines what counts as a correct retrieval. In transformed-memory stores that is exactly the contested object, yet it is typically left implicit. This is a benchmark-design problem rather than a system-tuning one: unlike hyperparameter sensitivity analysis, which varies a system knob while holding evaluation fixed, our audit varies the evaluation definition while holding system output fixed. It is also distinct from classical IR evaluation-sensitivity analysis, which varies relevance judgments or changes the retrieval unit globally rather than crediting system-generated descendants of a single source turn; we develop this contrast in Section \ref{sec:related}.

We study this setting on LoCoMo and LongMemEval-S across four native
retrievers. Switching only the scoring target changes Raw versus
Canonical nDCG on 83.4\% to 94.0\% of shared queries. Transfer runs
with Mem0 and MemoryOS remain target sensitive but show different target
orderings, confirming that the effect is a benchmark validity problem on
fixed ranked outputs rather than an artifact of target size alone.

% \begin{figure}[t]
% \centering
% \includegraphics[width=1\columnwidth]{figures/fig_target_noninvariance.pdf}
% \caption{Target non-invariance in retrieval evaluation. A query retrieves a derived memory (rank 1) that is factually correct and traces back to the gold raw turn, yet Raw scoring records a miss because the retrieved item lacks a raw-turn ID. Source and Canonical scoring both credit the hit.}
% \label{fig:qual_example}
% \end{figure}

This paper makes three contributions.
\begin{itemize}
    \setlength\itemsep{2pt}
    \setlength\parskip{0pt}
    \setlength\parsep{0pt}

    \item We identify \emph{target noninvariance}, a benchmark validity
    problem in transformed conversational memory: when ranked outputs
    are fixed, changing only the admissible target set can change
    scores, system rankings, and store design recommendations.

    \item We show across two primary benchmarks, one supplementary
    stress test, and two external memory architectures that the effect
    is large, affecting 83.4\% to 94.0\% of shared queries, surviving
    transfer to Mem0 and MemoryOS, and not being removed by post-hoc
    aggregation.

    \item We provide TIAP, a fixed-output audit procedure, and
    MTEL-Mem, a reproducibility layer that packages target mappings,
    source fixtures, and rescoring utilities for auditing saved traces.
\end{itemize}

We do not argue that any one target is universally correct. Rather, when
a benchmark can credit different stored forms of the same evidence,
scoring target choice is part of the evaluation design and should be
defined and reported explicitly.

\section{Related Work}
\label{sec:related}

Conversational-memory benchmarks such as LoCoMo, LongMemEval-S, and
BEAM have established retrieval- and QA-based evaluation for long-horizon
dialogue systems \citep{locomo, longmemeval, beam}. They differ in
answer-level scoring: LoCoMo uses lexical metrics such as F1 and BLEU-1
alongside retrieval recall over annotated evidence turns, while
LongMemEval-S uses a GPT-4o judge for flexible answer correctness and
reports Recall and nDCG when retrieval traces expose answer-location
labels. Our audit is upstream of these choices. It asks which stored
memory IDs are eligible to satisfy the annotated evidence before either
a lexical scorer or a judge evaluates the final answer.

Memory-system papers \citep{generative-agents, memorybank, memgpt,
mem0, memoryos, zep, theanine, rmm, memwalker, readagent} focus on
architecture, storage policy, and memory management. Many systems
extract, summarize, or canonicalize conversational evidence before
serving it. We study the evaluation target induced by these storage
choices: when raw turns and derived memories are linked to the same
evidence, which stored IDs should receive credit?

The closest analogy is classical IR work on evaluation sensitivity
\citep{bailey-judges, yang-neural-hype, craswell-dltrack,
voorhees-philosophy, voorhees-harman, manning-ir,
jarvelin-kekalainen}. Prior work shows that conclusions can be fragile
under incomplete judgments \citep{buckley-voorhees, zobel-assessor,
sanderson-zobel, saxton-assessor}, and document-versus-passage
evaluation shows that the scored unit can matter even when the evidence
is related \citep{yang-neural-hype, craswell-dltrack}. BEIR further
shows that retrieval conclusions depend on benchmark design choices
\citep{beir}. However, these settings usually vary judgments over fixed
units or change the retrieval unit globally. In transformed memory
stores, the system itself creates multiple source-linked descendants of
the same evidence inside one evaluated index, and the benchmark must
decide which descendants are admissible targets.

RAG evaluation frameworks such as RAGAs and ARES evaluate whether
retrieved evidence supports an answer \citep{ragas, ares}, but they
hold the candidate evidence unit fixed. The same ambiguity can arise in
chunk-based RAG, abstractive summarization indices, and
knowledge-graph-augmented retrieval whenever an index stores
non-identity transformations of source documents. Our audit makes this
question explicit for conversational memory: before answer-level
evaluation begins, which stored descendants of the same source evidence
count as admissible retrieval targets?

\section{Method}
\label{sec:method}

\begin{figure*}[!ht]
    \centering
    \includegraphics[width=0.98\textwidth]{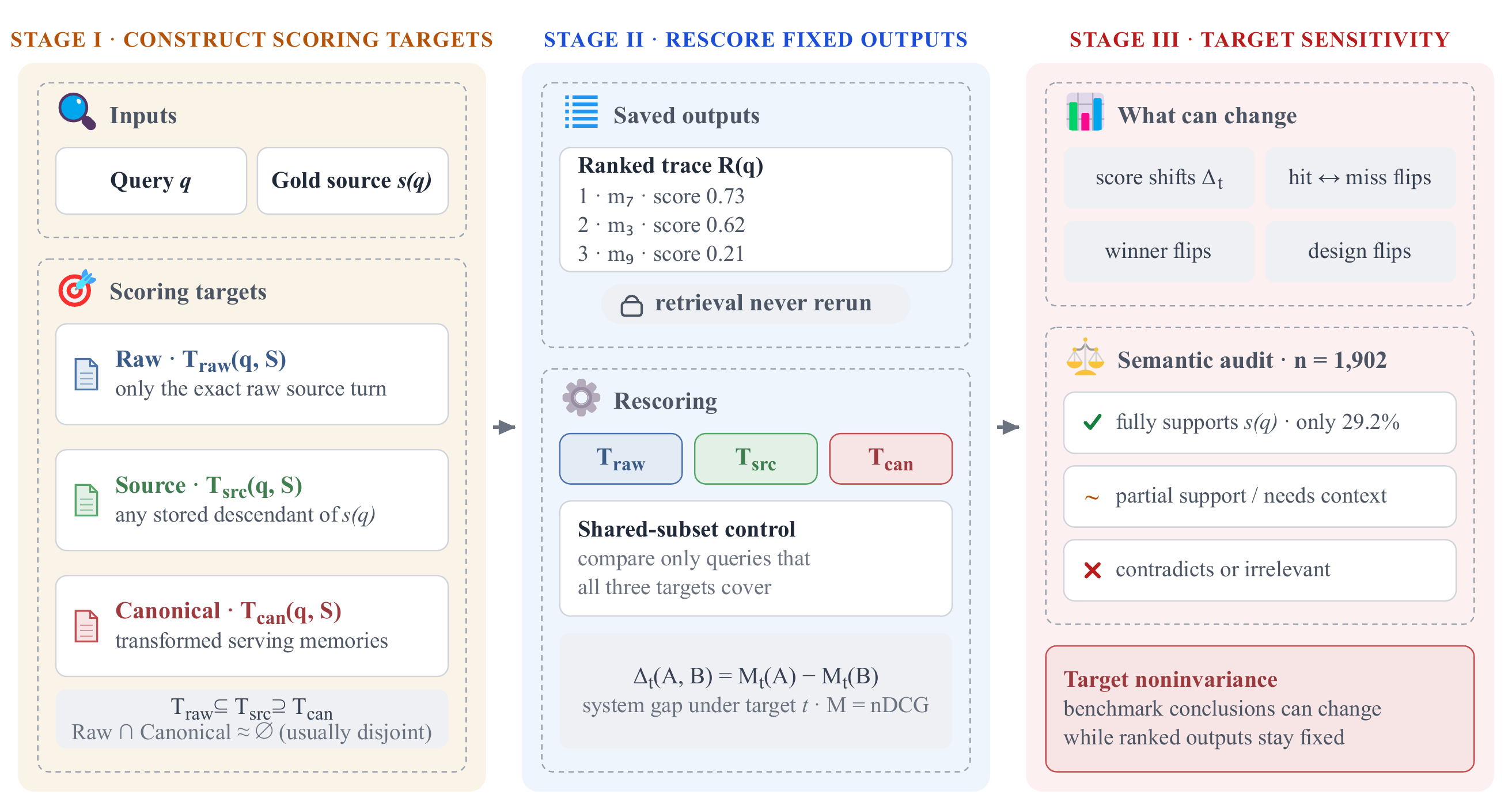}
    \vspace{-3 mm}
    \caption{
    Overview of the TIAP audit framework. Stage I constructs Raw,
    Source, and Canonical scoring targets from the same query, memory
    store, and source evidence. Stage II performs fixed output rescoring
    by applying the three targets to the same saved ranked traces and
    controlling for shared query coverage. Stage III analyzes target
    sensitivity and audits contested credits for semantic support.
    }
        \vspace{-3 mm}
    \label{fig:tiap_overview}
\end{figure*}

Figure~\ref{fig:tiap_overview} summarizes the TIAP audit framework in
three stages. First, TIAP constructs Raw, Source, and Canonical scoring
targets for each query and memory store. It then rescores the same
saved ranked traces under these targets without rerunning retrieval.
Finally, it measures whether the credited target changes scores,
winners, design recommendations, or the semantic validity of contested
credits.
\vspace{-2 mm}
\subsection{Stage I: Constructing Scoring Targets}
\label{sec:targets}

\textbf{Notation.}
Let $q$ be a benchmark query, $S$ a memory store,
$\mathcal{I}(S)$ the set of stored memory IDs in $S$, and
$R_k^S(q)$ the top $k$ ranked output emitted by $S$. Let $s(q)$ denote
the source evidence that anchors the gold label. For any stored memory
$m \in \mathcal{I}(S)$, let $\ell_S(m)$ be its source anchor. We use
$\mathrm{raw}_S(m)$ to indicate that $m$ is the raw source turn and
$\mathrm{can}_S(m)$ to indicate that $m$ is a transformed canonical
memory intended for serving.

\textbf{Scoring targets.}
A \emph{scoring target} is the set of stored memory IDs eligible to
receive credit for a query. We evaluate three targets:
\rawtxt{Raw}, which credits only the exact raw source turn;
\sourcetxt{Source}, which credits any stored descendant linked to the
same source evidence; and \cantxt{Canonical}, which credits transformed
non-raw descendants intended for serving. Canonical is a serving-form
target set rather than a single globally canonical record, since a
source may have multiple transformed descendants.

Formally, for a fixed source anchor $s(q)$ and store $S$:
\vspace{-3 mm}
\begingroup\small
\begin{align}
T_{\mathrm{raw}}(q,S)
&= \{m \in \mathcal{I}(S) : \ell_S(m)=s(q),
    \mathrm{raw}_S(m)\}, \\
T_{\mathrm{src}}(q,S)
&= \{m \in \mathcal{I}(S) : \ell_S(m)=s(q)\}, \\
T_{\mathrm{can}}(q,S)
&= \{m \in \mathcal{I}(S) : \ell_S(m)=s(q),
    \mathrm{can}_S(m)\}.
\end{align}
\endgroup

By construction, $T_{\mathrm{raw}}(q,S) \subseteq
T_{\mathrm{src}}(q,S)$ and $T_{\mathrm{can}}(q,S) \subseteq
T_{\mathrm{src}}(q,S)$. In most transformed memory stores,
$T_{\mathrm{raw}}(q,S)$ and $T_{\mathrm{can}}(q,S)$ are disjoint:
the former contains the original evidence turn, while the latter
contains transformed serving memories derived from it. We call the
rules a store uses to assign $\ell_S$, $\mathrm{raw}_S$, and
$\mathrm{can}_S$ its \emph{lineage contract}. When this contract is
shared across compared systems, we omit $S$ for readability.

\subsection{Stage II: Fixed Output Rescoring}
\label{sec:rescoring}

\textbf{Audit question.}
TIAP asks what changes when only the scoring target changes. It answers
this by rescoring the same saved ranked outputs under \rawtxt{Raw},
\sourcetxt{Source}, and \cantxt{Canonical} targets. Retrieval is not
rerun; the ranked traces remain fixed, and only the credited memory IDs
change.
% memory IDs changes. \textcolor{red}{Appendix~\ref{app:tiap_algorithm} gives the complete implementable audit procedure.}

\textbf{System gap under a target.}
For systems $A$ and $B$, metric $M$, and target
$t \in \{\mathrm{raw}, \mathrm{src}, \mathrm{can}\}$, we define
\begin{equation}
    \Delta_t(A,B) = M_t(A) - M_t(B),
\end{equation}
where $M_t$ is $M$ evaluated against target set $T_t(q,S)$. If the sign
of $\Delta_t(A,B)$ changes as $t$ varies, the benchmark conclusion is
not target-invariant

This differs from standard assessor disagreement, where items are fixed
and only judgments vary. Here, the credited IDs themselves change with
the store: a benchmark using only \rawtxt{Raw} credit has already
decided that a raw turn can succeed while a canonical descendant of the
same evidence cannot. TIAP makes this representational assumption
explicit and tests whether conclusions survive it.

\textbf{Shared query subset control.}
Target coverage can differ across \rawtxt{Raw}, \sourcetxt{Source},
and \cantxt{Canonical} scoring because not every source turn produces a
transformed serving memory. Directly comparing targets on different
query sets would confound scoring target choice with query difficulty.
TIAP therefore repeats target comparisons on shared query subsets when
coverage differs. A query is included only when all compared targets
have at least one eligible stored ID, ensuring that score changes
reflect the credited target set rather than changes in the evaluated
query population.

\subsection{Stage III: Target Sensitivity and Semantic Audit}
\label{sec:sensitivity}

\textbf{Target sensitivity outputs.}
After rescoring, TIAP records score shifts, query-level hit status
changes, winner flips, and design recommendation changes. A winner flip
occurs when the sign of $\Delta_t(A,B)$ changes across scoring targets,
so the preferred system depends on which stored representation is
credited.
A design recommendation is the memory store or configuration that
appears best under a chosen metric and scoring target. If two
configurations are ranked differently under \rawtxt{Raw},
\sourcetxt{Source}, and \cantxt{Canonical} while their saved ranked
outputs remain fixed, then the benchmark recommendation depends on the
scoring target rather than on retrieval behavior.

\textbf{Semantic audit of contested credits.}
A broader target can award more retrieval credit, but source lineage
alone does not guarantee semantic support: a transformed memory may omit
the answer, remove a relevant condition, change the temporal scope, or
preserve only weakly related context.
We therefore audit contested credits where \rawtxt{Raw} misses but
\sourcetxt{Source} or \cantxt{Canonical} receives credit. Each credited
memory is judged against the source evidence $s(q)$ and the answer
required by $q$, then assigned one of three labels: \emph{supports} if
it fully supports the answer, \emph{partial} if it contains related
evidence but requires additional context, and \emph{does not support} if
it contradicts, omits, or is irrelevant to the required evidence. This
separates provenance from semantic support and measures whether relaxed
credit is justified rather than merely available.

\section{Experimental Setup}
\label{sec:setup}

\subsection{Datasets}
\label{sec:datasets}

We instantiate the audit on two conversational memory benchmarks,
LoCoMo \textcolor{blue}{\citep{locomo}} and LongMemEval-S
\textcolor{blue}{\citep{longmemeval}}. We also use BEAM \citep{beam} as a
supplementary stress test. We select these benchmarks because each
annotates gold evidence at the level of source conversation turns or
sessions. This makes source linked lineage recoverable and allows us to
construct the Raw, Source, and Canonical target distinction.
Table~\ref{tab:dataset_stats} summarizes the coverage used by the
audit.
\textbf{LoCoMo} \textcolor{blue}{\citep{locomo} } evaluates long term
conversational memory and annotates turn IDs for QA gold answers. It
reports lexical automatic metrics for answer quality and retrieval
recall over annotated evidence turns. This makes it suitable for testing
whether different stored forms of the same source turn receive different
retrieval credit.
\textbf{LongMemEval-S} \textcolor{red}{\citep{longmemeval}} benchmarks chat
assistants on long term interactive memory. It annotates evidence
sessions and positions within them, uses a judge based answer scorer,
and reports Recall and nDCG when retrieval traces expose answer location
labels. Its evidence annotations allow us to recover the source anchor
for each query and construct target mappings over stored memories.
\textbf{BEAM} \citep{beam} provides a grounded million token slice with
a \texttt{source\_chat\_ids} field. We use only its 1M grounded slice
and treat it as a single provider robustness check rather than part of
the main native run matrix.
Across all benchmarks, queries, reference answers, and evidence labels
are held out and used only at evaluation time. They are never provided
to the parser that produces transformed memories, so they cannot leak
into the memory store.
\begin{table}[t]
\centering
\footnotesize
\caption{Dataset coverage used by the audit. BEAM-1M is supplementary.}
\vspace{-3 mm}
\label{tab:dataset_stats}
{\setlength{\tabcolsep}{2pt}
\begin{tabularx}{\columnwidth}{>{\raggedright\arraybackslash}p{0.27\columnwidth}>{\raggedleft\arraybackslash}p{0.14\columnwidth}>{\raggedleft\arraybackslash}p{0.11\columnwidth}>{\raggedleft\arraybackslash}p{0.22\columnwidth}>{\raggedleft\arraybackslash}p{0.19\columnwidth}}
\toprule
\theadrow
Dataset & Fixture & Eval & Native runs & Canon.\ cov. \\
\midrule
LoCoMo & 5,882 & 1,977 & 1,510--1,533 & 899--963 \\
LongMemEval-S & 231,595 & 470 & 439--470 & 299--332 \\
BEAM-1M (supp.) & 74,630 & 625 & 625 & 408 \\
\bottomrule
\end{tabularx}
}
\vspace{-2.5 mm}
\end{table}

\begin{table*}[t]
\centering
\small
\caption{
Fixed-subset rescoring on canonical-covered shared subsets for all
eight native runs. Canonical exceeds Raw on every shared subset; Source
exceeds Canonical on all four LoCoMo shared subsets.
}
\label{tab:shared_subset}
\vspace{-2mm}
{\setlength{\tabcolsep}{3pt}
\renewcommand{\arraystretch}{0.98}
\begin{tabular}{llcccccc}
\toprule
\theadrow
Dataset & Provider & $n$ & \rawcell{Raw} & \sourcecell{Source} &
\canonicalcell{Canonical} & Canonical--Raw & Canonical--Source \\
\midrule
\multirow{4}{*}{LoCoMo}
& Lexical & 899 & 0.1876 & 0.3809 & 0.3556 &
+0.168 [0.147, 0.189] & -0.025 [-0.036, -0.015] \\
& all MiniLM & 899 & 0.2118 & 0.4198 & 0.3875 &
+0.176 [0.154, 0.198] & -0.032 [-0.043, -0.022] \\
& BGE M3 & 963 & 0.2548 & 0.4109 & 0.3231 &
+0.068 [0.045, 0.091] & -0.088 [-0.101, -0.075] \\
& mxbai-embed-large & 899 & 0.2120 & 0.4195 & 0.3870 &
+0.175 [0.152, 0.198] & -0.033 [-0.043, -0.022] \\
\midrule
\multirow{4}{*}{LongMemEval-S}
& Lexical & 332 & 0.1877 & 0.4041 & 0.4558 &
+0.268 [0.236, 0.303] & +0.052 [0.038, 0.066] \\
& all MiniLM & 332 & 0.2001 & 0.4561 & 0.5309 &
+0.331 [0.299, 0.364] & +0.075 [0.061, 0.089] \\
& BGE M3 & 299 & 0.1969 & 0.4576 & 0.5326 &
+0.336 [0.302, 0.371] & +0.075 [0.061, 0.090] \\
& mxbai-embed-large & 332 & 0.1815 & 0.5032 & 0.6015 &
+0.420 [0.382, 0.457] & +0.098 [0.083, 0.113] \\
\bottomrule
\end{tabular}
}
\vspace{-2mm}
\end{table*}

\subsection{Target Mappings}
\label{sec:target_mappings}
% Each transformed memory stores the source identifier from which it was
% produced, so target mappings are recovered from explicit lineage
% metadata rather than from text similarity.
For each query, we map the benchmark source evidence to Raw, Source,
and Canonical targets using source linked lineage metadata. The mapping
is deterministic: benchmark source fixtures are joined to the raw memory
ID and to all transformed descendant IDs exported with the same source
anchor. This allows every saved ranked trace to be rescored under the
three target definitions without rerunning retrieval.
The primary native store contains the raw source turn together with
parser derived transformed memories linked back to the same source
evidence. Source unions the raw and transformed IDs associated with a
source fixture. Canonical keeps only non raw transformed IDs, so it is a
serving form target set rather than a guarantee of one canonical item
per source. The parser operates only on conversation turns.
Canonical coverage can differ from Raw coverage because not every
source turn produces a transformed serving memory. When a query has no
Canonical target, we exclude it from Canonical comparisons. Raw and
Source comparisons use their available shared query sets. This makes
shared subset rescoring part of the main analysis rather than an
optional check.
For the parser density sweep, $D_1$, $D_5$, and $D_8$ denote
\emph{parser\_max\_facts} equal to 1, 5, and 8 transformed facts per
source turn.
Target-set sizes are small in practice. Across the 13{,}620
query-system pairs in the native runs, Source and Raw contain the same
IDs for 44.3\% of pairs, Source adds exactly one ID for 32.3\%, and two
or more IDs for the remaining 23.4\%.

% \vspace{-3 mm}
\subsection{Retrieval and Memory Systems}
\label{sec:retrieval_setup}

We evaluate four native retrievers: lexical search, all-MiniLM-L6-v2
\citep{sentence-bert}, BGE-M3 \citep{bge-m3}, and
mxbai-embed-large \citep{mxbai-embed}. All main results are computed
from saved top $k=60$ ranked traces and rescored under the target sets
defined in Section~\ref{sec:targets}. We use $k=60$ because it matches
or exceeds the retrieval depth used by the memory systems in our scope
and provides enough depth for source-linked descendants to appear in the
ranked output. On saved traces, Recall@$k$ plateaus after approximately
$k=40$, with less than six percentage points of additional recall from
$k=40$ to $k=60$ across retriever and dataset combinations. Using
$k=60$ therefore reduces the risk that target sensitivity is caused by
an unusually shallow retrieval cutoff, while keeping the audit focused
on evidence that would plausibly be available to a deployed memory
system. Rescoring the same saved traces at $k=20$ and $k=40$
strengthens the effect: the relative Source minus Raw shift grows from
$+57\%$ at $k=60$ to $+84\%$ at $k=20$ on LoCoMo, and from $+123\%$ to
$+215\%$ on LongMemEval-S
(Appendix~\ref{app:k_sensitivity}). The reported $k=60$ results are
therefore the most conservative depth for the effect.
All native runs use extractive answer mode and random seed
\texttt{1337}. The lexical run uses the same benchmark harness without
dense embeddings, and embedding providers are served locally via Ollama.
Shared-subset paired statistics use 3{,}000 paired bootstrap resamples,
percentile confidence intervals, and RNG seed \texttt{1337}. The
transfer analysis includes full LoCoMo and LongMemEval-S runs for both
Mem0 \citep{mem0} and MemoryOS \citep{memoryos}. Zep \citep{zep} is
out of scope because its graph-native memories do not expose a stable
source-turn to descendant ID mapping.

\subsection{Semantic Audit Methodology}
\label{sec:audit_method}
\vspace{-1 mm}
The semantic audit covers all 1{,}902
\texttt{raw\_miss / source\_hit / canonical\_hit} disagreement cases
extracted from the eight native runs. These cases are selected by a
deterministic extraction script over saved traces and constitute the
complete set of native run cases where Raw misses while both Source and
Canonical receive credit. Labels follow the three-way rubric defined in
Section~\ref{sec:sensitivity}: \texttt{supports}, \texttt{partial}, and
\texttt{does\_not\_support}. Auxiliary serving semantics tags are
reported only as exploratory notes and are not used as primary evidence.

\begin{table}[!t]
\centering
\small
\caption{Inter-rater agreement on the 115-case validation subset used to confirm rubric consistency before scaling to the full 1{,}902-case audit. Binary labels collapse \emph{supports} $\cup$ \emph{partial} $\to$ \emph{relevant}. Fleiss' $\kappa$ is computed over all five raters; pairwise values are Cohen's $\kappa$.}
\vspace{-2.5 mm}
\label{tab:llm_agreement}
{\setlength{\tabcolsep}{3.5pt}
\begin{tabular}{lcccc}
\toprule
\theadrow
 & \multicolumn{2}{c}{Binary} & \multicolumn{2}{c}{Three-class} \\
\cmidrule(lr){2-3}\cmidrule(lr){4-5}
\theadrow
Rater / pair & $\kappa$ & Agree & $\kappa$ & Agree \\
\midrule
\multicolumn{5}{l}{\textit{All five raters (Fleiss)}} \\
\quad 1 human + 4 LLMs & \textbf{0.83} & n/a & 0.79 & n/a \\
\midrule
\multicolumn{5}{l}{\textit{Each model vs.\ human (Cohen)}} \\
\quad GPT-5.5 & 0.87 & 93.9\% & 0.83 & 88.7\% \\
\quad Claude Opus 4.7 & 0.85 & 93.0\% & 0.83 & 88.7\% \\
\quad Gemini 3.1 Pro & 0.82 & 91.3\% & 0.74 & 82.6\% \\
\quad DeepSeek V4 Pro & 0.77 & 88.7\% & 0.73 & 81.7\% \\
\midrule
\multicolumn{5}{l}{\textit{Model--model pairs (Cohen, binary)}} \\
\quad Gemini $\leftrightarrow$ GPT & 0.91 & 95.7\% & n/a & n/a \\
\quad Claude $\leftrightarrow$ GPT & 0.91 & 95.7\% & n/a & n/a \\
\quad Claude $\leftrightarrow$ Gemini & 0.89 & 94.8\% & n/a & n/a \\
\quad Claude $\leftrightarrow$ DeepSeek & 0.77 & 88.7\% & n/a & n/a \\
\quad GPT $\leftrightarrow$ DeepSeek & 0.75 & 87.8\% & n/a & n/a \\
\quad Gemini $\leftrightarrow$ DeepSeek & 0.74 & 87.0\% & n/a & n/a \\
\bottomrule
\end{tabular}
}
\end{table}

To validate the rubric before scaling, we first had a human annotator
label a 115-case subset drawn from LongMemEval-S with all-MiniLM and
LoCoMo with BGE-M3. The subset was stratified by credited rank bucket,
using ranks 1 to 5, 6 to 20, and 21 to 60, and evenly spaced by query
index within each bucket. A second human reviewer independently labeled
a 36-case overlap subset, and disagreements were adjudicated. We then
ran four frontier LLMs on the same 115 cases found at {Table~\ref{tab:llm_agreement}}. Across the five raters,
the validation subset achieved Fleiss' $\kappa = 0.83$ under a binary
mapping, indicating that the rubric produced consistent classifications
on this subset.
After this validation, we applied the rubric to all 1{,}902 contested
cases using five frontier LLMs via the OpenRouter API: Claude Sonnet
4, Claude Opus 4.7, Gemini 3.1 Pro, GPT-5.5, and DeepSeek V4 Pro. Each
model judged each case independently at temperature~0.
% Across the 1{,}876 cases where all five models returned valid labels, mean pairwise binary agreement is 90.9\% and mean three-class agreement is 85.9\% (Table~\ref{tab:full_audit_agreement}). We report per-model label distributions and a five-model majority vote in Table~\ref{tab:semantic_audit}.

\section{Results}
\label{sec:results}

We organize the results around whether scoring targets change scores,
store design recommendations, query-type behavior, relaxed-credit
reliability, or transfer behavior.

\subsection{Scoring Target Alters Retrieval Scores}
\label{sec:score_shift}

The central result is that target choice materially changes retrieval
scores even when ranked outputs are unchanged. On canonical-covered
shared subsets, switching only the scoring target changes nDCG on
83.4\% to 94.0\% of \rawtxt{Raw} versus \cantxt{Canonical} rescoring
comparisons across LoCoMo and LongMemEval-S. Thus, target choice is not
a minor implementation detail: it changes the measured relevance of the
same ranked outputs for most evaluated queries.

Coverage must be controlled explicitly. Queries with a Canonical target
are easier under Raw retrieval than queries without Canonical coverage:
Raw nDCG is higher by a mean run-level gap of $+0.086$ with bootstrap
95\% CI $[+0.049,+0.123]$, while answer F1 changes much less, with a
mean gap of $+0.008$ and CI $[+0.002,+0.014]$. We therefore report
Canonical versus Raw comparisons only on canonical-covered shared query
subsets.
The shared subset is not skewed toward simple queries. Covered queries
are slightly easier under Raw at the query level (mean Raw nDCG 0.206
versus 0.167 for excluded queries) but contain proportionally more
Multi-hop questions (22.4\% versus 8.1\%), while excluded queries are
dominated by Single-hop questions (58.2\% versus 38.9\%). Because
target sensitivity remains high across all major query categories
(Section~\ref{sec:query_breadth}), the composition of the shared
subset does not explain the instability we report.

The score shifts are large and target dependent
(Table~\ref{tab:shared_subset}). On LongMemEval-S,
\cantxt{Canonical} exceeds \rawtxt{Raw} for every provider. The largest
Canonical minus Raw effects appear for BGE-M3, with $+0.336$ and CI
$[0.302,0.371]$, and mxbai-embed-large, with $+0.420$ and CI
$[0.382,0.457]$. On LoCoMo, \cantxt{Canonical} also exceeds
\rawtxt{Raw} for every provider, but \sourcetxt{Source} remains above
\cantxt{Canonical} for all four providers. The strongest Source versus
Canonical reversal appears on BGE-M3, where Canonical minus Source is
$-0.088$ and the confidence interval excludes zero.
Table~\ref{tab:query_instability_compact} reports the compact
query-level instability matrix behind the headline result. The rate of
nonzero nDCG change ranges from 83.4\% to 94.0\% across the eight native
runs. The shifts are not concentrated near zero: in the full
distribution, 65\% of LoCoMo and 71\% of LongMemEval-S query-provider
pairs shift by at least $0.1$ nDCG
(Figure~\ref{fig:ndcg_shift_dist} in Appendix~\ref{app:pairwise}).
Thus, target choice changes both absolute retrieval scores and the
relative evaluation landscape across datasets and retrievers.
\begin{table}[t]
\centering
\small
\caption{
\rawtxt{Raw} versus \cantxt{Canonical} query-level nDCG instability.
}
\vspace{-1.8 mm}
\label{tab:query_instability_compact}
\setlength{\tabcolsep}{4pt}
\begin{tabular}{llrrr}
\toprule
Dataset & Provider & $|Q|$ & $\Delta\neq 0$ & Rate \\
\midrule
LoCoMo & Lexical    & 899 & 758 & 84.3\% \\
LoCoMo & all MiniLM & 899 & 794 & 88.3\% \\
LoCoMo & BGE M3     & 963 & 851 & 88.4\% \\
LoCoMo & mxbai      & 899 & 794 & 88.3\% \\
\midrule
LongMemEval-S & Lexical    & 332 & 277 & 83.4\% \\
LongMemEval-S & all MiniLM & 332 & 309 & 93.1\% \\
LongMemEval-S & BGE M3     & 299 & 281 & 94.0\% \\
LongMemEval-S & mxbai      & 332 & 312 & 94.0\% \\
\bottomrule
\end{tabular}
\end{table}

\subsection{Score Shifts Are Not Explained by Target-Set Size}
\label{sec:cardinality}

A possible confound is that relaxed targets simply contain more
eligible IDs. Cardinality contributes to the effect but does not
explain it. Across the 13{,}620 query-system pairs, the correlation
between the target-set size increase ($|$Source$|-|$Raw$|$) and the
Source minus Raw nDCG shift is $r=0.38$ ($p<10^{-15}$), so set size
explains about 14\% of the shift variance. Queries where Source adds
exactly one eligible ID ($n=4{,}399$) already shift by $+0.179$ mean
nDCG ($t=46.8$, $p<10^{-10}$), and the shift saturates after two to
three extra IDs rather than growing with set size
(Appendix~\ref{app:cardinality}).
For the Raw versus Canonical comparison, the confound is weaker still.
On the canonical-covered subset ($n=7{,}413$), set size explains at
most 1.7\% of the Canonical minus Raw shift variance, and a single
extra eligible ID yields a $+0.153$ mean shift ($t=27.7$). The
instability is therefore representational: it depends on which stored
forms receive credit, not on how many.

\subsection{Scoring Target Changes Store Design Recommendations}
\label{sec:design_flip}

The score shifts also change an engineering decision. We run a parser
density sweep on a fixed matched subset with $n=453$, using
\texttt{parser\_max\_facts} values of 1, 5, and 8. We denote these
settings as $D_1$, $D_5$, and $D_8$. Changing only the scoring target
changes which density appears best.
For the lexical comparison between $D_1$ and $D_5$, \rawtxt{Raw}
prefers the sparser $D_1$ store, \sourcetxt{Source} prefers the denser
$D_5$ store, and \cantxt{Canonical} returns to $D_1$. A practitioner
reading only the Raw-scored benchmark would choose
\texttt{parser\_max\_facts} equal to 1, saving extraction cost, storage,
and index size. The same saved comparison under Source scoring instead
recommends \texttt{parser\_max\_facts} equal to 5. The ranked outputs
are fixed across these rescorings. What changes is which source-linked
memory IDs are allowed to receive credit.

Across the six pairwise comparisons in Table~\ref{tab:density_flip1},
\sourcetxt{Source} prefers the denser setting in all six comparisons,
\rawtxt{Raw} prefers the sparsest setting in four of six comparisons,
and \cantxt{Canonical} is split. Thus, target choice can affect
downstream engineering decisions, not merely benchmark reporting.
%%%%%%%%%%%%%%%%%%%%%%%%
\begin{table}[!t]
\centering
\small
\caption{Target-dependent winner by parser density ($D_1, D_5, D_8$) on the fixed shared subset ($n=453$). The winning configuration depends on whether scoring uses \rawtxt{Raw}, \sourcetxt{Source}, or \cantxt{Canonical}.}
\vspace{-3 mm}
\label{tab:density_flip1}
\begin{tabular}{l l ccc}
\toprule
\theadrow
\textbf{Provider} & \textbf{Comp.} & \rawcell{\textbf{Raw}} & \sourcecell{\textbf{Source}} & \canonicalcell{\textbf{Canonical}} \\
\midrule
Lexical & $D_1$ vs $D_5$ & \rawcell{$D_1$} & \sourcecell{$D_5$} & \canonicalcell{$D_1$} \\
& $D_1$ vs $D_8$ & \rawcell{$D_1$} & \sourcecell{$D_8$} & \canonicalcell{$D_1$} \\
& $D_5$ vs $D_8$ & \rawcell{$D_5$} & \sourcecell{$D_8$} & \canonicalcell{$D_8$} \\
\midrule
all-MiniLM & $D_1$ vs $D_5$ & \rawcell{$D_1$} & \sourcecell{$D_5$} & \canonicalcell{$D_1$} \\
& $D_1$ vs $D_8$ & \rawcell{$D_1$} & \sourcecell{$D_8$} & \canonicalcell{$D_1$} \\
& $D_5$ vs $D_8$ & \rawcell{$D_8$} & \sourcecell{$D_8$} & \canonicalcell{$D_8$} \\
\bottomrule
\end{tabular}
\end{table}

%%%%%%%%%%%%%%%%%%%%%%%
\begin{table*}[t]
\centering

\begin{minipage}[t]{0.49\textwidth}
\centering
\scriptsize
\setlength{\tabcolsep}{2.5pt}
\renewcommand{\arraystretch}{1.05}
\captionof{table}{
Semantic audit of contested credits. Majority denotes the five-model
majority-vote label.
}
\vspace{-3 mm}
\label{tab:semantic_audit}
\begin{tabular}{l|lrrrr}
\toprule
\rowcolor{gray!20}
Dataset & Judge & $n$ & Supports & Partial &  Not support\\
\midrule
\multirow{6}{*}{\rotatebox{90}{LoCoMo}}& Sonnet 4 & 1{,}278 & 384 (30.0\%) & 553 (43.3\%) & 341 (26.7\%) \\
& Opus 4.7 & 1{,}278 & 409 (32.0\%) & 513 (40.1\%) & 356 (27.9\%) \\
& Gemini 3.1 & 1{,}278 & 447 (35.0\%) & 487 (38.1\%) & 344 (26.9\%) \\
& GPT-5.5 & 1{,}269 & 413 (32.5\%) & 588 (46.3\%) & 268 (21.1\%) \\
& DeepSeek V4 & 1{,}264 & 409 (32.4\%) & 577 (45.6\%) & 278 (22.0\%) \\
\cmidrule(lr){2-6}
& \textbf{Majority} & \textbf{1{,}256} & \textbf{416 (33.1\%)} &
\textbf{514 (40.9\%)} & \textbf{326 (26.0\%)} \\
\midrule
\multirow{6}{*}{\rotatebox{90}{LongMemEval-S}}& Sonnet 4 & 624 & 120 (19.2\%) & 206 (33.0\%) & 298 (47.8\%) \\
& Opus 4.7 & 624 & 122 (19.6\%) & 203 (32.5\%) & 299 (47.9\%) \\
& Gemini 3.1 & 624 & 136 (21.8\%) & 227 (36.4\%) & 261 (41.8\%) \\
& GPT-5.5 & 622 & 147 (23.6\%) & 238 (38.3\%) & 237 (38.1\%) \\
& DeepSeek V4 & 622 & 135 (21.7\%) & 218 (35.0\%) & 269 (43.2\%) \\
\cmidrule(lr){2-6}
& \textbf{Majority} & \textbf{620} & \textbf{132 (21.3\%)} &
\textbf{229 (36.9\%)} & \textbf{259 (41.8\%)} \\
\bottomrule
\end{tabular}
\end{minipage}
\hfill
\begin{minipage}[t]{0.49\textwidth}
\centering
\scriptsize
\setlength{\tabcolsep}{3pt}
\renewcommand{\arraystretch}{1.08}
\captionof{table}{
Transfer validation on Mem0 and MemoryOS under Recall@60, MRR, and
nDCG@60.
}
\vspace{-3 mm}
\label{tab:transfer}
\begin{tabular}{>{\centering\arraybackslash}p{0.08\linewidth}|llrrrr}
\toprule
\rowcolor{gray!20}
System & Dataset & Target & $|Q|$ & R@60 & MRR & nDCG@60 \\
\midrule
\multirow{6}{*}{\rotatebox{90}{Mem0}}
& \multirow{3}{*}{LoCoMo}& \rawcell{Raw} & 1977 & 0.723 & 0.158 & 0.275 \\
& & \sourcecell{Source} & 1977 & 0.945 & 0.620 & 0.663 \\
& & \canonicalcell{Canonical} & 1977 & 0.901 & 0.577 & 0.617 \\
\cmidrule(lr){2-7}
& \multirow{3}{*}{LongMemEval-S}
& \rawcell{Raw} & 470 & 0.978 & 0.454 & 0.567 \\
& & \sourcecell{Source} & 470 & 0.987 & 0.687 & 0.717 \\
& & \canonicalcell{Canonical} & 470 & 0.758 & 0.412 & 0.435 \\
\midrule
\multirow{6}{*}{\rotatebox{90}{MemoryOS}}
& \multirow{3}{*}{LoCoMo}& \rawcell{Raw} & 1977 & 0.770 & 0.364 & 0.434 \\
& & \sourcecell{Source} & 1977 & 0.994 & 0.481 & 0.579 \\
& & \canonicalcell{Canonical} & 1977 & 0.990 & 0.265 & 0.409 \\
\cmidrule(lr){2-7}
& \multirow{3}{*}{LongMemEval-S}
& \rawcell{Raw} & 470 & 0.359 & 0.207 & 0.231 \\
& & \sourcecell{Source} & 470 & 0.995 & 0.789 & 0.812 \\
& & \canonicalcell{Canonical} & 470 & 0.989 & 0.682 & 0.730 \\
\bottomrule
\end{tabular}
\end{minipage}
\vspace{-4 mm}
\end{table*}

\subsection{Target Sensitivity Is Broad Across Query Types}
\label{sec:query_breadth}

Target sensitivity is not confined to one query family. Across shared
queries, Raw versus Canonical rescoring changes nDCG for 83.6\% to
93.2\% of queries within major LoCoMo categories and for 77.1\% to
100.0\% of queries within LongMemEval-S categories. The effect also
persists in categories with lower Canonical coverage, including LoCoMo
Single-hop and Temporal questions and LongMemEval-S single-session
questions.

A useful boundary appears in Source versus Canonical comparisons on
LongMemEval-S. In that benchmark, instability is concentrated in
multi-session and knowledge-update questions, while it is near zero for
single-session assistant questions. This suggests that Source versus
Canonical target choice matters less when the answer is a simple
single-session fact, but becomes more consequential when a query
requires aggregation, update tracking, or reasoning across sessions.

\subsection{Relaxed Credit Is Only Partially Reliable}
\label{sec:semantic_audit_results}

A natural concern is that relaxed targets may give credit too easily.
Using the semantic audit methodology in Section~\ref{sec:audit_method},
we audit all 1{,}902
\texttt{raw\_miss / source\_hit / canonical\_hit} disagreement cases
across the eight native runs.

Five frontier LLMs were prompted to judge every case independently
(Table~\ref{tab:semantic_audit}). Under the five-model majority vote,
29.2\% of contested credits fully support the query, 39.6\% are
partially supportive, and 31.2\% do not support it. The gap between
datasets is substantial. On LoCoMo, 33.1\% of relaxed credit is fully
justified, compared with 21.3\% on LongMemEval-S. The non-supportive
rate is 26.0\% on LoCoMo and 41.8\% on LongMemEval-S.
The judge agreement statistics show the same qualitative pattern.
Across the 1{,}876 cases where all five models returned valid labels,
mean pairwise binary agreement is 90.9\%, and mean three-class
agreement is 85.9\% (Appendix \ref{app:full_audit_agreement} Table~\ref{tab:full_audit_agreement}). Even under
the most permissive binary collapse, where \texttt{supports} and
\texttt{partial} are both treated as relevant, 31.2\% of relaxed credit
is entirely unjustified. The partial category, which accounts for
39.6\% of cases, requires additional context that may or may not appear
elsewhere in the ranked list. Therefore, provenance-relaxed credit
cannot be treated as automatically valid semantic evidence.

Target choice also propagates to answer-level metrics. On LoCoMo, Raw
best predicts answer F1, while on LongMemEval-S, Source and Canonical
align better with the judge-based scorer
(Appendix~\ref{app:answer_alignment}). Raw-only hits on LoCoMo carry a
mean answer F1 of 0.123 versus 0.046 for Canonical-only hits, so the
credited target changes measured answer quality rather than being
absorbed by the reader. This confirms that no single target is
universally correct even at the downstream answer level, and that
end-to-end answer scoring does not remove the need to specify the
retrieval target: answer metrics conflate retrieval quality with
reader compensation, and none of the 11 papers in our reporting audit
specifies which stored descendant receives retrieval credit
(Appendix~\ref{app:reporting_audit}).
% This result explains how to interpret the fixed output score changes. Target sensitivity is real, but the semantic correctness of relaxed credit is mixed. \cantxt{Canonical} should be reported alongside \rawtxt{Raw} rather than treated as a replacement. Relaxed credit is informative, but it is not reliable enough to stand alone.

% \input{figures/tab_semantic_audit}

\subsection{Target Sensitivity Transfers Across Memory Architectures}
\label{sec:transfer_results}

To test whether the effect extends beyond the primary parser-derived
store, we evaluate Mem0~\citep{mem0} and MemoryOS~\citep{memoryos}.
The transfer runs remain strongly target-sensitive across Recall@60,
MRR, and nDCG@60, but the target ordering is not uniform
(Table~\ref{tab:transfer}). On full LoCoMo runs, Mem0 follows the same
ordering across all three metrics, with \sourcetxt{Source} above
\cantxt{Canonical} and \cantxt{Canonical} above \rawtxt{Raw}. MemoryOS
also places \sourcetxt{Source} highest on LoCoMo, but under the
rank-sensitive metrics MRR and nDCG@60 it shows the nonmonotone pattern
\sourcetxt{Source} above \rawtxt{Raw} and \rawtxt{Raw} above
\cantxt{Canonical}. On full LongMemEval-S runs, MemoryOS is monotone
across all three metrics, whereas Mem0 is nonmonotone, with
\sourcetxt{Source} above \rawtxt{Raw} and \rawtxt{Raw} above
\cantxt{Canonical}. Unlike the native parser runs, these transfer
systems have zero targetless queries for all three targets, so full run
values are also shared subset values.
These runs rebut a simple inflation story. A more abstract target is
not always better, because \rawtxt{Raw} and \cantxt{Canonical} trade
places depending on where the relevant detail survives and which metric
is used. The effect is therefore not confined to one store design.
%

% \begin{table}[t]
% \centering
% \scriptsize
% \setlength{\tabcolsep}{4pt}
% \caption{External transfer validation on Mem0 and MemoryOS.}
% \vspace{-3mm}
% \label{tab:transfer_compact}
% \begin{tabular}{p{0.38\linewidth}p{0.21\linewidth}p{0.1\linewidth}p{0.17\linewidth}}
% \toprule
% System and data & Target & $Q$ & nDCG@60 \\
% \midrule
% Mem0 LoCoMo & \rawtxt{Raw} & 1977 & 0.275 \\
% Mem0 LoCoMo & \sourcetxt{Source} & 1977 & 0.663 \\
% Mem0 LoCoMo & \cantxt{Canonical} & 1977 & 0.617 \\
% \midrule
% Mem0 LongMemEval-S & \rawtxt{Raw} & 470 & 0.567 \\
% Mem0 LongMemEval-S & \sourcetxt{Source} & 470 & 0.717 \\
% Mem0 LongMemEval-S & \cantxt{Canonical} & 470 & 0.435 \\
% \midrule
% MemoryOS LoCoMo & \rawtxt{Raw} & 1977 & 0.434 \\
% MemoryOS LoCoMo & \sourcetxt{Source} & 1977 & 0.579 \\
% MemoryOS LoCoMo & \cantxt{Canonical} & 1977 & 0.409 \\
% \midrule
% MemoryOS LongMemEval-S & \rawtxt{Raw} & 470 & 0.231 \\
% MemoryOS LongMemEval-S & \sourcetxt{Source} & 470 & 0.812 \\
% MemoryOS LongMemEval-S & \cantxt{Canonical} & 470 & 0.730 \\
% \bottomrule
% \end{tabular}
% \end{table}
%
As a further stress test on a different retrieval unit and grounding
structure, we rescore BEAM's grounded 1M slice~\citep{beam} under one
provider. On the full 625 query run, nDCG@60 moves from $0.055$ under
\rawtxt{Raw} to $0.097$ under \sourcetxt{Source} and $0.197$ under
\cantxt{Canonical}. On the shared subset with $n=408$, the ordering
persists, with $0.071 < 0.134 < 0.197$.

\subsection{Choosing a Scoring Target}
\label{sec:target_guidance}

The appropriate target depends on what the deployed system serves
(Table~\ref{tab:target_guidance}). \cantxt{Canonical} matches
deployments that serve transformed memories, with \rawtxt{Raw}
reported alongside it as a stricter provenance floor. \rawtxt{Raw} is
primary when users consume original turns or when exact provenance
matters. \sourcetxt{Source} is a provenance-relaxed stress test rather
than a default headline target: only 29.2\% of its contested credits
fully support the query
(Section~\ref{sec:semantic_audit_results}). Whichever target is
chosen, comparisons should use shared subsets when coverage differs,
and winner flips across targets should be reported explicitly.

\begin{table}[!t]
\centering
\small
\caption{Deployment-oriented target selection.}
\label{tab:target_guidance}
\begin{tabular}{p{0.42\columnwidth} l p{0.24\columnwidth}}
\toprule
\textbf{What users consume} & \textbf{Primary} & \textbf{Co-report} \\
\midrule
Transformed serving memories & \cantxt{Canonical} & \rawtxt{Raw} as floor \\
Original conversation turns & \rawtxt{Raw} & \cantxt{Canonical} if served \\
Any source-linked evidence & \sourcetxt{Source} & semantic audit \\
\bottomrule
\end{tabular}
\end{table}

\section{Conclusion}
\label{sec:conclusion}

Conversational memory benchmarks often evaluate stores that contain both
raw dialogue turns and transformed descendants of the same evidence. We
show that the credited scoring target is not a minor detail: under fixed
ranked outputs, changing only this target alters scores on 83\% to
94\% of shared queries, changes store design recommendations, and
transfers across memory architectures.

The semantic audit further shows that relaxed credit is only partially
reliable, since source-linked descendants do not always preserve the
evidence needed to answer the query. Benchmark reports should therefore
specify the credited target set, use shared-subset rescoring when
coverage differs, and flag cases where target choice changes the
winner. TIAP makes this sensitivity visible and reproducible.

\section{Limitations}
This study is about transformed conversational memory, not IR evaluation in general.
The native headline matrix is concentrated in one parser-derived store construction.
The Mem0 and MemoryOS runs broaden the evidence, but they are still a small architectural sample.

{The semantic audit covers all 1{,}902 disagreement cases across the eight native runs using five frontier LLM judges whose rubric was validated on a 115-case human-annotated subset (Fleiss' $\kappa = 0.83$, 90.9\% mean pairwise binary agreement on the full set).
The five-model majority vote reduces single-model bias, but the human validation subset is still small (115 cases, one primary annotator with a 36-case adjudicated overlap), so prevalence estimates should be interpreted accordingly.}

Canonical coverage is not random.
Canonical-covered queries have higher Raw nDCG than uncovered queries in the native runs, so full-set Raw scores are not directly comparable to Canonical scores on the covered subset.
This is why the main Raw--Canonical claims use matched canonical-covered query subsets.

TIAP currently treats target membership as binary once a retrieved item falls inside a target set.
It does not assign graded credit to partially sufficient evidence or compose support across multiple retrieved facts.
This hurts most for queries whose answer is distributed across evidence units: multi-hop questions, multi-session preference questions, temporal comparisons, counts, totals, and other aggregation queries.
In those cases, a retrieved item can be provenance-correct but insufficient, or one of several necessary facts.
Our semantic audit surfaces exactly this boundary, so binary target membership should be read as an audit baseline rather than a complete model of evidential sufficiency.

The Raw / Source / Canonical basis used here is a minimal audit basis for the source-linked transformed-memory systems we study.
It is useful because these systems expose recoverable lineage from transformed memories back to benchmark source evidence.
Other memory architectures may expose different stored units and may require different target sets or audit constructions.
We therefore treat this basis as an instantiation for this setting, not as a complete taxonomy of conversational memory.
The audit also assumes stable source anchors $s(q)$, recoverable lineage from retrieved IDs back to those anchors, enumerable target sets over the descendants, and exported fixed ranked traces.
Parser- or extraction-based stores satisfy this directly, which is why our current instantiation covers the native store, Mem0, and MemoryOS.
Graph-native memories may require node-, edge-, or path-level target definitions beyond the Raw/Source/Canonical basis used here.

More broadly, TIAP and \mtelmem{} are meant to make target sensitivity auditable, not to prescribe a universal target set or a single pass/fail threshold.
Extending the audit to graph-native memories and other retrieval settings remains future work.

\FloatBarrier
\bibliography{references}
\clearpage
\appendix

\section*{Supplementary Material}
\addcontentsline{toc}{section}{Supplementary Material}
\section{Query-Level Instability Details}
\label{app:pairwise}

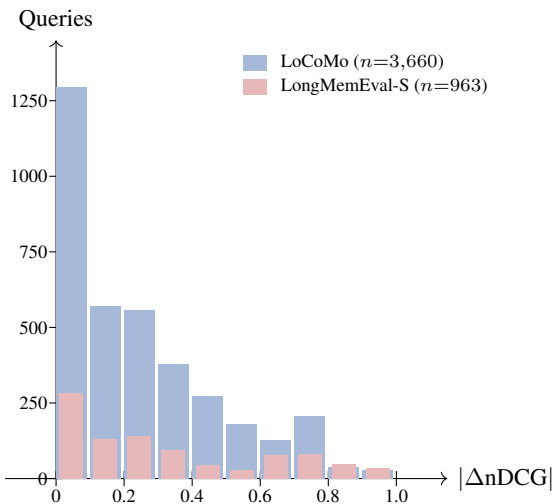
\begin{figure}[!b]
\centering
\begin{tikzpicture}[x=4.5cm, y=0.004cm]
  % Axes
  \draw[->] (-0.15,0) -- (1.15,0) node[right,font=\small] {$|\Delta\text{nDCG}|$};
  \draw[->] (0,0) -- (0,1450) node[above,font=\small] {Queries};

  % Y-axis ticks
  \foreach \y/\l in {0/0, 250/250, 500/500, 750/750, 1000/1000, 1250/1250} {
    \draw (-0.02,\y) -- (0.02,\y) node[left,font=\scriptsize,xshift=-2pt] {\l};
  }

  % X-axis ticks
  \foreach \x/\l in {0/0, 0.2/0.2, 0.4/0.4, 0.6/0.6, 0.8/0.8, 1.0/1.0} {
    \draw (\x,-25) -- (\x,25) node[below,font=\scriptsize,yshift=-3pt] {\l};
  }

  % LoCoMo bars (real data: n=3660, 4 providers x shared queries)
  % bins: [1294, 572, 557, 380, 272, 181, 129, 208, 38, 29]
  \fill[RawTurnColor!50] (0,0) rectangle (0.09,1294);
  \fill[RawTurnColor!50] (0.1,0) rectangle (0.19,572);
  \fill[RawTurnColor!50] (0.2,0) rectangle (0.29,557);
  \fill[RawTurnColor!50] (0.3,0) rectangle (0.39,380);
  \fill[RawTurnColor!50] (0.4,0) rectangle (0.49,272);
  \fill[RawTurnColor!50] (0.5,0) rectangle (0.59,181);
  \fill[RawTurnColor!50] (0.6,0) rectangle (0.69,129);
  \fill[RawTurnColor!50] (0.7,0) rectangle (0.79,208);
  \fill[RawTurnColor!50] (0.8,0) rectangle (0.89,38);
  \fill[RawTurnColor!50] (0.9,0) rectangle (0.99,29);

  % LongMemEval-S bars overlay (real data: n=963, 3 providers x shared queries)
  % bins: [282, 131, 142, 93, 44, 29, 77, 80, 49, 36]
  \fill[CanonicalColor!40] (0.01,0) rectangle (0.08,282);
  \fill[CanonicalColor!40] (0.11,0) rectangle (0.18,131);
  \fill[CanonicalColor!40] (0.21,0) rectangle (0.28,142);
  \fill[CanonicalColor!40] (0.31,0) rectangle (0.38,93);
  \fill[CanonicalColor!40] (0.41,0) rectangle (0.48,44);
  \fill[CanonicalColor!40] (0.51,0) rectangle (0.58,29);
  \fill[CanonicalColor!40] (0.61,0) rectangle (0.68,77);
  \fill[CanonicalColor!40] (0.71,0) rectangle (0.78,80);
  \fill[CanonicalColor!40] (0.81,0) rectangle (0.88,49);
  \fill[CanonicalColor!40] (0.91,0) rectangle (0.98,36);

  % Legend
  \fill[RawTurnColor!50] (0.55,1350) rectangle (0.62,1400);
  \node[right,font=\scriptsize] at (0.63,1375) {LoCoMo ($n{=}3{,}660$)};
  \fill[CanonicalColor!40] (0.55,1270) rectangle (0.62,1320);
  \node[right,font=\scriptsize] at (0.63,1295) {LongMemEval-S ($n{=}963$)};
\end{tikzpicture}
\caption{Per-query $|\Delta\text{nDCG}|$ when switching from \rawtxt{Raw} to \cantxt{Canonical} targets, pooled across native providers on shared canonical-covered subsets. 65\% of LoCoMo and 71\% of LongMemEval-S query--provider pairs shift by ${\geq}\,0.1$ nDCG (mean $|\Delta|$: 0.25 and 0.33, respectively).}
\label{fig:ndcg_shift_dist}
\end{figure}

Figure~\ref{fig:ndcg_shift_dist} shows the per-query magnitude of
nDCG change when switching from \rawtxt{Raw} to \cantxt{Canonical}
targets. The distribution is not concentrated near zero: 65\% of LoCoMo
and 71\% of LongMemEval-S query-provider pairs shift by at least 0.1
nDCG, with mean $|\Delta|$ of 0.25 and 0.33, respectively.
Table~\ref{tab:pairwise} expands the compact instability summary in
Table~\ref{tab:query_instability_compact} to the full set of pairwise
target comparisons. For each of the eight provider--dataset pairs, it
reports the number of shared queries, the count of binary hit flips at
$k{=}60$, top-1 rank flips, the number of queries whose nDCG changed,
and the overall change rate. \rawtxt{Raw} versus \cantxt{Canonical}
comparisons show the highest change rates (83.4\%--94.0\%), while
\rawtxt{Raw} versus \sourcetxt{Source} rates are lower but still
substantial (31.3\%--66.2\%). \sourcetxt{Source} versus
\cantxt{Canonical} rates cluster between 44.2\% and 55.2\%, confirming
that the effect is not driven solely by the \rawtxt{Raw}--\cantxt{Canonical} gap.

\section{Aggregation and Mitigation Checks}
\label{app:mitigation}

A natural follow-up is whether a single aggregate score can absorb
target sensitivity. We test three aggregation functions over the
per-query nDCG triplet
$(\mathrm{nDCG}_{\rawtxt{Raw}},
\mathrm{nDCG}_{\sourcetxt{Source}},
\mathrm{nDCG}_{\cantxt{Canonical}})$: arithmetic mean, geometric mean, and per-query minimum.
Table~\ref{tab:mitigation} reports per-provider nDCG@60 under the three
aggregation functions. No aggregation produces a ranking distinct from
an existing individual target. On LoCoMo, all three aggregations
reproduce the Raw or Source provider ordering. On LongMemEval-S, the
aggregation function determines which ordering emerges: geometric mean
and minimum reproduce Raw, while arithmetic mean reproduces Source or
Canonical. Aggregation therefore selects one target-dependent ordering
rather than removing the target choice.
Agreement filtering also fails. Restricting evaluation to queries where
all three targets agree on binary hit status retains 72\% to 85\% of
queries, but nDCG values still diverge, with mean per-query spreads
between 0.16 and 0.30. Provider rankings on LongMemEval-S continue to
flip between targets, showing that agreement on binary retrieval does
not imply agreement on graded relevance.
\begin{table}[!t]
\centering
\small
\caption{Provider nDCG@60 under individual targets and three ensemble functions across all eight native runs. No ensemble produces a novel consensus ranking. $\tau_d$: Kendall-tau distance from the Raw provider ordering.}
\vspace{-2 mm}
\label{tab:mitigation}
\resizebox{\columnwidth}{!}{%
\begin{tabular}{@{}ll cccc c@{}}
\toprule
\theadrow
\textbf{Dataset} & \textbf{Method} & \textbf{lexical} & \textbf{allminilm} & \textbf{bge\_m3} & \textbf{mxbai} & \textbf{$\tau_d$} \\
\midrule
\multirow{6}{*}{LoCoMo}
& Raw              & .176 & .198 & .240 & .197 & n/a  \\
& Source           & .290 & .320 & .339 & .318 & 0 \\
& Canonical        & .275 & .301 & .283 & .299 & 2 \\
\cmidrule(lr){2-7}
& Arith.\ mean    & .247 & .273 & .288 & .272 & 0 \\
& Geom.\ mean     & .202 & .228 & .240 & .227 & 0 \\
& Min              & .148 & .166 & .175 & .165 & 0 \\
\midrule
\multirow{6}{*}{\shortstack[l]{LongMem-\\Eval-S}}
& Raw              & .154 & .157 & .155 & .134 & n/a \\
& Source           & .307 & .338 & .333 & .361 & 3 \\
& Canonical        & .344 & .391 & .384 & .430 & 3 \\
\cmidrule(lr){2-7}
& Arith.\ mean    & .268 & .295 & .291 & .309 & 3 \\
& Geom.\ mean     & .215 & .228 & .226 & .187 & 0 \\
& Min              & .147 & .154 & .153 & .133 & 0 \\
\bottomrule
\end{tabular}%
}
\vspace{-2 mm}
\end{table}

\begin{table}[!t]
\centering
\small
\caption{Supplementary BEAM-1M lexical stress test. The ordering remains \rawtxt{Raw} $<$ \sourcetxt{Source} $<$ \cantxt{Canonical} both on the full run and on the canonical-covered shared subset.}
\vspace{-2 mm}
\label{tab:beam_stress}
\resizebox{\linewidth}{!}{%
\setlength{\tabcolsep}{4pt}
\begin{tabular}{lcccc}
\toprule
\theadrow
Slice & $Q$ & \rawcell{Raw} & \sourcecell{Source} & \canonicalcell{Canonical} \\
\midrule
Full run & 625 & 0.0554 & 0.0969 & 0.1969 \\
Shared subset & 408 & 0.0706 & 0.1342 & 0.1969 \\
\bottomrule
\end{tabular}
}
\end{table}

\section{Supplementary BEAM Runs}
\label{app:beam}

The main paper uses BEAM as a supplementary stress test. This appendix
provides the supporting BEAM details, which are not shown in the main
tables. On the full 625-query run, nDCG@60
moves from 0.055 under \rawtxt{Raw} to 0.097 under
\sourcetxt{Source} and 0.197 under \cantxt{Canonical}. On the
canonical-covered shared subset with $n=408$, the same ordering
persists. Table~\ref{tab:beam_support} reports supporting BEAM-100K runs. Both
the lexical and mxbai settings preserve the same
\rawtxt{Raw} $<$ \sourcetxt{Source} $<$ \cantxt{Canonical} ordering on
the full run and on the canonical-covered shared subset.

\begin{table}[!t]
\centering
\scriptsize
\caption{Supporting BEAM-100K runs with consistent
\rawtxt{Raw} $<$ \sourcetxt{Source} $<$ \cantxt{Canonical} ordering.}
\vspace{-2.5mm}
\label{tab:beam_support}
{\setlength{\tabcolsep}{4pt}
\begin{tabular}{llcccc}
\toprule
\theadrow
Scale & Setting & $Q$ & \rawcell{Raw} & \sourcecell{Source} & \canonicalcell{Canonical} \\
\midrule
100K & Lexical full & 355 & 0.1105 & 0.2064 & 0.3489 \\
100K & Lexical shared subset & 246 & 0.1430 & 0.2815 & 0.3489 \\
100K & mxbai full & 355 & 0.1233 & 0.2281 & 0.3898 \\
100K & mxbai shared subset & 245 & 0.1617 & 0.3135 & 0.3898 \\
\bottomrule
\end{tabular}
}
\end{table}

\section{Qualitative Disagreement Cases}
\label{app:qual_cases}

Table~\ref{tab:qual_cases} lists representative disagreement cases from
LongMemEval-S with all-MiniLM. These examples illustrate how the same
saved ranked output can receive different scores depending on whether
credit is assigned to the raw source turn, a source-linked descendant,
or a canonical serving memory.
\begin{table}[!t]
\centering
\scriptsize
\caption{Representative LongMemEval-S all-MiniLM disagreement cases. Scores are nDCG.}
\vspace{-3 mm}
\label{tab:qual_cases}
{\setlength{\tabcolsep}{2pt}
\begin{tabularx}{\columnwidth}{>{\raggedright\arraybackslash}Xccc}
\toprule
\theadrow
Query & \rawcell{\textbf{Raw}} & \sourcecell{\textbf{Src.}} & \canonicalcell{\textbf{Can.}} \\
\midrule
What degree did I graduate with? & \rawcell{0.000} & \sourcecell{0.613} & \canonicalcell{1.000} \\
How long is my daily commute to work? & \rawcell{0.173} & \sourcecell{1.000} & \canonicalcell{1.000} \\
What play did I attend at the local community theater? & \rawcell{0.000} & \sourcecell{0.500} & \canonicalcell{0.500} \\
How long did I wait for the decision on my asylum application? & \rawcell{---} & \sourcecell{0.171} & \canonicalcell{0.000} \\
\bottomrule
\end{tabularx}
}
\end{table}

\section{Auxiliary Answer-Level Alignment}
\label{app:answer_alignment}

This appendix provides additional detail for the answer-alignment
analysis referenced in Section~\ref{sec:semantic_audit_results}. 
Table~\ref{tab:qa_disagreement} reports the disagreement-cell counts and mean answer F1.
We
join target-specific hit labels to answer F1 on the same saved traces.
On LoCoMo, \rawtxt{Raw} has the strongest average answer-alignment
signal across the eight native runs. On LongMemEval-S,
\sourcetxt{Source} and \cantxt{Canonical} align better, though on a
thinner base of four native runs.

This difference is consistent with the benchmarks' answer scorers:
LoCoMo's lexical scoring rewards exact answer wording more directly,
whereas LongMemEval-S's judge-based scoring is more tolerant of
paraphrased or transformed evidence. This analysis does not replace the
paper's main result. Instead, it provides a downstream check showing
that answer-level alignment does not identify one universally preferred
target set.

\begin{table}[h!]
\centering
\small
\caption{Secondary query-level answer-alignment check on shared \rawtxt{Raw}-versus-\cantxt{Canonical} eligible queries. `\rawtxt{Raw only}' and `\cantxt{Canonical only}' denote disagreement cells where hit@60 changes under identical ranked outputs. Means are averaged over native full runs within each dataset; counts are summed across those runs.}
\vspace{-2 mm}
\label{tab:qa_disagreement}
\setlength{\tabcolsep}{3pt}
\resizebox{\columnwidth}{!}{
\begin{tabular}{lrrrrrr}
\toprule
\theadrow
Dataset & \rawcell{Raw-only $n$} & \canonicalcell{Canon-only $n$} & \rawcell{Raw-only F1} & \canonicalcell{Canon-only F1} & \rawcell{Raw-only strong} & \canonicalcell{Canon-only strong} \\
\midrule
LoCoMo & \rawcell{507} & \canonicalcell{1142} & \rawcell{0.123} & \canonicalcell{0.046} & \rawcell{0.063} & \canonicalcell{0.013} \\
LongMemEval-S & \rawcell{32} & \canonicalcell{312} & \rawcell{0.030} & \canonicalcell{0.065} & \rawcell{0.000} & \canonicalcell{0.004} \\
\bottomrule
\end{tabular}
}
\end{table}

\section{Semantic Audit Agreement}
\label{app:llm_agreement}

To validate the semantic audit rubric before scaling, we ran four
frontier LLMs with extended reasoning on a 115-case validation subset:
GPT-5.5, Claude Opus 4.7, Gemini 3.1 Pro, and DeepSeek V4 Pro. These labels
were compared with the human validation labels. Each model received the
same rubric and case material and returned one of
\{\emph{supports}, \emph{partial}, \emph{does\_not\_support}\}. The
goal is to test rubric reproducibility rather than model accuracy.
Table~\ref{tab:llm_agreement} (in the main text) reports Fleiss' $\kappa$ across the five
raters and pairwise Cohen's $\kappa$ for both the original three-class
and collapsed binary schemes. All four models reach substantial to
almost perfect consistency with the human labels.

\section{Full-Audit Multi-Model Agreement}
\label{app:full_audit_agreement}

After validating the rubric, we ran all five LLMs on the complete
1{,}902-case audit. Of these cases, 1{,}876 received valid labels from
all five models; 26 cases returned empty responses from GPT-5.5 or
DeepSeek V4 Pro after retries. Table~\ref{tab:full_audit_agreement}
reports pairwise agreement on the 1{,}876 common-valid cases.

Mean binary agreement across all ten model pairs is 90.9\%, and mean
three-class agreement is 85.9\%. The five-model majority vote matches
the original Sonnet 4 labels on 90.1\% of cases, reducing concern that
the semantic audit is driven by a single model. The majority-vote
distribution is 29.2\% supports, 39.6\% partial, and 31.2\% does not
support. The dataset-level gap, approximately 33\% fully supportive on
LoCoMo versus 21\% on LongMemEval-S, is reproduced by all five judges.
{\color{blue}
\begin{table}[t]
\centering
\small
\caption{Pairwise agreement among five LLM judges on the full 1{,}902-case semantic audit ($n=1{,}876$ cases where all five returned valid labels). Binary collapses \emph{supports}$\,\cup\,$\emph{partial}$\,\to\,$\emph{relevant}.}
\vspace{-2.5 mm}
\label{tab:full_audit_agreement}
{\setlength{\tabcolsep}{3.5pt}
\begin{tabular}{lcc}
\toprule
\theadrow
Model pair & 3-class agree & Binary agree \\
\midrule
Sonnet 4 \;$\leftrightarrow$\; Opus 4.7     & 85.9\% & 90.0\% \\
Sonnet 4 \;$\leftrightarrow$\; Gemini 3.1   & 85.8\% & 91.3\% \\
Sonnet 4 \;$\leftrightarrow$\; GPT-5.5      & 84.7\% & 89.4\% \\
Sonnet 4 \;$\leftrightarrow$\; DeepSeek V4  & 83.9\% & 88.4\% \\
Opus 4.7 \;$\leftrightarrow$\; Gemini 3.1   & 89.9\% & 94.2\% \\
Opus 4.7 \;$\leftrightarrow$\; GPT-5.5      & 86.1\% & 91.3\% \\
Opus 4.7 \;$\leftrightarrow$\; DeepSeek V4  & 83.4\% & 88.9\% \\
Gemini 3.1 \;$\leftrightarrow$\; GPT-5.5    & 89.9\% & 94.1\% \\
Gemini 3.1 \;$\leftrightarrow$\; DeepSeek V4 & 83.6\% & 90.0\% \\
GPT-5.5 \;$\leftrightarrow$\; DeepSeek V4   & 86.4\% & 90.9\% \\
\midrule
Mean (all 10 pairs)                          & 85.9\% & 90.9\% \\
\bottomrule
\end{tabular}
}
\end{table}
}

\section{Reporting-Practice Audit}
\label{app:reporting_audit}

Table~\ref{tab:reporting_audit} audits whether 11 recent
conversational-memory papers publicly specify the components needed for
a scoring-target ontology. We inspect only what each paper states or
releases; we do not claim the authors lacked internal conventions.

\paragraph{Scope.}
We selected papers cited in our study that either propose a
conversational-memory benchmark or propose a memory system evaluated on
retrieval or QA tasks. For each paper, we checked four questions:

\begin{enumerate}
    \item \textbf{Transforms memory?} Does the system or benchmark
    produce non-raw representations, such as extracted facts, summaries,
    or knowledge-graph entries, from raw conversation turns?
    \item \textbf{Reports retrieval unit?} Is the scored unit formally
    defined?
    \item \textbf{Reports target mapping?} Is there a documented
    mapping from scored items to source conversation evidence?
    \item \textbf{Specifies descendant credit?} When one source turn
    produces multiple stored items, does the paper specify which
    descendants receive credit?
\end{enumerate}
\providecommand{\ymark}{\textcolor{SourceFamilyColor}{$\bullet$}}
\providecommand{\pmark}{\textcolor{RawTurnColor}{$\circ$}}
\providecommand{\nmark}{\textcolor{CanonicalColor}{$\times$}}

\begin{table}[!t]
\centering
\scriptsize
\setlength{\tabcolsep}{3.5pt}
\renewcommand{\arraystretch}{1.02}
\caption{
Reporting-practice audit of target specification in 11 conversational-memory papers. All transform memory; none fully specifies which stored descendants receive scoring credit.
\ymark{}\,{=}\,explicit, \pmark{}\,{=}\,partial,
\nmark{}\,{=}\,not specified.
}
\vspace{-2 mm}
\label{tab:reporting_audit}
\begin{tabular}{@{}lcccc@{}}
\toprule
\theadrow
\textbf{Paper}
  & \textbf{Trans.}
  & \textbf{Unit}
  & \textbf{Map}
  & \textbf{Credit} \\
\midrule
\multicolumn{5}{@{}l}{\textit{Benchmarks}} \\
LoCoMo \citep{locomo}              & \ymark & \ymark & \pmark & \nmark \\
LongMemEval-S \citep{longmemeval}  & \ymark & \pmark & \pmark & \nmark \\
BEAM \citep{beam}                  & \ymark & \pmark & \pmark & \nmark \\
\midrule
\multicolumn{5}{@{}l}{\textit{Memory systems}} \\
Mem0 \citep{mem0}                  & \ymark & \pmark & \nmark & \nmark \\
MemoryOS \citep{memoryos}          & \ymark & \pmark & \nmark & \nmark \\
Zep \citep{zep}                    & \ymark & \pmark & \nmark & \nmark \\
Theanine \citep{theanine}          & \ymark & \pmark & \nmark & \nmark \\
RMM \citep{rmm}                    & \ymark & \pmark & \nmark & \nmark \\
MemGPT \citep{memgpt}              & \ymark & \pmark & \nmark & \nmark \\
MemoryBank \citep{memorybank}      & \ymark & \pmark & \nmark & \nmark \\
LD-Agent \citep{helloagain}        & \ymark & \pmark & \nmark & \nmark \\
\bottomrule
\end{tabular}
\vspace{-2mm}
\end{table}
\paragraph{Findings.}
All 11 papers create transformed memory representations. Retrieval
units are described at the architectural level in most papers, but only
LoCoMo explicitly labels and separately evaluates distinct unit types.
No paper fully specifies a target mapping that traces credit through the
transformation layer to source evidence. No paper addresses credit
assignment when a single source turn produces multiple stored
descendants.

Three benchmarks provide partial provenance. LoCoMo annotates turn IDs
on QA gold answers~\citep{locomo}. LongMemEval-S annotates evidence
sessions and positions within them~\citep{longmemeval}. BEAM includes a
\texttt{source\_chat\_ids} field in its question schema, although some
published examples leave this field empty~\citep{beam}.
Two system papers touch on the gap indirectly. \citet{theanine} observe
that existing long-term conversation datasets lack gold mappings between
dialogue contexts and memories. \citet{zep} maintain episodic edges
linking extracted facts to source episodes but state that these
connections are not examined in their experiments.
This audit motivates our recommendation: transformed-memory evaluations
should state the scored target set and release target mappings alongside
retrieval traces.

\begin{table*}[!t]
\centering
\footnotesize
\caption{Query-level instability between target definitions across all eight main-matrix provider-dataset pairs.}
\label{tab:pairwise}
{\setlength{\tabcolsep}{2pt}
\begin{tabular}{llcccccc}
\toprule
\theadrow
Dataset & Provider & Pair & Queries & Hit flips & Top-1 flips & nDCG changed & Rate \\
\midrule
\multirow{12}{*}{LoCoMo} & \multirow{3}{*}{Lexical} & \pairRS & 1,533 & 181 & 173 & 526 & 34.3\% \\
& & \pairRC & 899 & 244 & 193 & 758 & 84.3\% \\
& & \pairSC & 899 & 63 & 20 & 397 & 44.2\% \\
\cmidrule(lr){2-8}
& \multirow{3}{*}{all-MiniLM} & \pairRS & 1,533 & 162 & 198 & 541 & 35.3\% \\
& & \pairRC & 899 & 226 & 217 & 794 & 88.3\% \\
& & \pairSC & 899 & 64 & 19 & 425 & 47.3\% \\
\cmidrule(lr){2-8}
& \multirow{3}{*}{BGE-M3} & \pairRS & 1,510 & 134 & 167 & 473 & 31.3\% \\
& & \pairRC & 963 & 259 & 214 & 851 & 88.4\% \\
& & \pairSC & 963 & 125 & 47 & 532 & 55.2\% \\
\cmidrule(lr){2-8}
& \multirow{3}{*}{mxbai-embed-large} & \pairRS & 1,533 & 162 & 190 & 541 & 35.3\% \\
& & \pairRC & 899 & 226 & 208 & 794 & 88.3\% \\
& & \pairSC & 899 & 64 & 18 & 425 & 47.3\% \\
\midrule
\multirow{12}{*}{LongMemEval-S} & \multirow{3}{*}{Lexical} & \pairRS & 470 & 50 & 113 & 259 & 55.1\% \\
& & \pairRC & 332 & 68 & 113 & 277 & 83.4\% \\
& & \pairSC & 332 & 18 & 0 & 158 & 47.6\% \\
\cmidrule(lr){2-8}
& \multirow{3}{*}{all-MiniLM} & \pairRS & 470 & 70 & 129 & 301 & 64.0\% \\
& & \pairRC & 332 & 78 & 129 & 309 & 93.1\% \\
& & \pairSC & 332 & 8 & 0 & 158 & 47.6\% \\
\cmidrule(lr){2-8}
& \multirow{3}{*}{BGE-M3} & \pairRS & 439 & 62 & 112 & 276 & 62.9\% \\
& & \pairRC & 299 & 67 & 112 & 281 & 94.0\% \\
& & \pairSC & 299 & 5 & 0 & 144 & 48.2\% \\
\cmidrule(lr){2-8}
& \multirow{3}{*}{mxbai-embed-large} & \pairRS & 470 & 130 & 152 & 311 & 66.2\% \\
& & \pairRC & 332 & 131 & 152 & 312 & 94.0\% \\
& & \pairSC & 332 & 1 & 0 & 154 & 46.4\% \\
\bottomrule
\end{tabular}
}
\end{table*}

\section{Cardinality Analysis}
\label{app:cardinality}

This appendix details the target-set cardinality analysis summarized
in Section~\ref{sec:cardinality}. All analyses rescore the same saved
ranked traces. Table~\ref{tab:cardinality_bins} bins query-system
pairs by the number of extra eligible IDs
($|$Source$|-|$Raw$|$). Table~\ref{tab:cardinality_corr} reports the
corresponding correlations. The Canonical columns use the
canonical-covered subset ($n=7{,}413$); the Source-minus-Raw size
difference is an upper bound on the Canonical expansion since Source
is a superset of Canonical, so the 1.7\% variance figure is an upper
bound on the cardinality contribution.

\begin{table}[!ht]
\centering
\small
\caption{Mean nDCG shift by target-set size difference. Full-set
column: Source minus Raw across all 13{,}620 query-system pairs.
Covered columns: canonical-covered subset.}
\label{tab:cardinality_bins}
\begin{tabular}{lrrr}
\toprule
Extra IDs & Full set & \multicolumn{2}{c}{Canonical-covered} \\
 & Src$-$Raw & Src$-$Raw & Can$-$Raw \\
\midrule
0 & 0.000 & -- & -- \\
1 & $+0.179$ & $+0.186$ & $+0.153$ \\
2 & $+0.234$ & $+0.236$ & $+0.226$ \\
3 & $+0.288$ & $+0.289$ & $+0.302$ \\
$\geq 4$ & $+0.275$ & $+0.271$ & $+0.298$ \\
\bottomrule
\end{tabular}
\end{table}

\begin{table}[!ht]
\centering
\small
\caption{Correlation between target-set size difference and nDCG
shift.}
\label{tab:cardinality_corr}
\setlength{\tabcolsep}{4pt}
\begin{tabular}{llrr}
\toprule
Comparison & Subset & $r$ & $r^2$ \\
\midrule
Src$-$Raw & Full set ($n{=}13{,}620$) & 0.379 & 14.4\% \\
Src$-$Raw & Covered ($n{=}7{,}413$) & 0.104 & 1.1\% \\
Can$-$Raw & Covered ($n{=}7{,}413$) & 0.129 & 1.7\% \\
\bottomrule
\end{tabular}
\end{table}

The full-set bins contain $n=6{,}032$, $4{,}399$, $1{,}688$, $807$,
and $694$ pairs respectively; the covered bins contain $n=4{,}241$,
$1{,}672$, $806$, and $694$. The shift at one extra ID is significant
in both settings ($t=46.8$ full set; $t=27.7$ covered Canonical minus
Raw; both $p<10^{-10}$), and the shift saturates after two to three
extra IDs rather than growing with set size.

\section{Retrieval-Depth Sensitivity}
\label{app:k_sensitivity}

Table~\ref{tab:k_sensitivity} repeats the Source-versus-Raw rescoring
of the saved traces at $k=20$ and $k=40$ across the native retriever
configurations (eight on LoCoMo, three on LongMemEval-S). The relative
shift grows as $k$ decreases because higher-ranked positions receive
more nDCG weight at shallow depth. The $k=60$ setting used in the main
results is therefore the most conservative of the three depths.

\begin{table}[!ht]
\centering
\small
\caption{Mean Raw and Source nDCG at retrieval depths $k=20,40,60$.}
\label{tab:k_sensitivity}
\setlength{\tabcolsep}{3.5pt}
\begin{tabular}{llrrrr}
\toprule
Dataset & $k$ & Raw & Source & Shift & Rel. \\
\midrule
LoCoMo & 20 & 0.139 & 0.256 & $+0.117$ & $+84\%$ \\
LoCoMo & 40 & 0.179 & 0.294 & $+0.114$ & $+64\%$ \\
LoCoMo & 60 & 0.194 & 0.306 & $+0.112$ & $+57\%$ \\
\midrule
LongMemEval-S & 20 & 0.087 & 0.274 & $+0.187$ & $+215\%$ \\
LongMemEval-S & 40 & 0.102 & 0.293 & $+0.191$ & $+188\%$ \\
LongMemEval-S & 60 & 0.150 & 0.335 & $+0.185$ & $+123\%$ \\
\bottomrule
\end{tabular}
\end{table}

\section{Parser Operational Details}
\label{app:parser_details}

The native store parser converts conversation turns into transformed
memories. It operates only on conversation turns: queries, reference
answers, and evidence labels are held out and never provided to the
parser, so they cannot leak into the memory store. The
\emph{parser\_max\_facts} setting caps the number of transformed facts
extracted per source turn; the density sweep uses caps of 1, 5, and 8
($D_1$, $D_5$, $D_8$). Every transformed memory records the source
identifier of the turn it was derived from, and target mappings are
recovered from this lineage metadata rather than from text similarity.
Not every source turn yields a transformed serving memory, which is
why Canonical coverage differs from Raw coverage
(Section~\ref{sec:target_mappings}). All native runs use extractive
answer mode with random seed 1337, and embedding providers are served
locally via Ollama.

% \appendix

% \input{sections/12_appendix_query_level_instability}

% \input{sections/13_appendix_external_transfer_detail}

% \input{sections/14_appendix_beam_runs}

% \input{sections/15_appendix_density_example}

% \input{sections/16_appendix_qualitative_cases}

% \input{sections/17_appendix_answer_alignment}

% \input{sections/18_appendix_llm_agreement}

% \input{sections/19_appendix_mitigation}

% \input{sections/20_appendix_reporting_audit}

\end{document}